\documentclass{eas}
\usepackage{graphicx}
\usepackage{txfonts}
\usepackage{natbib}
\begin{document}

\title{Inferring properties of small convective cores in main-sequence solar-like pulsators} 
\author{I.\,M.\,Brand\~ao}\address{Centro de Astrof\'isica and Faculdade de Ci\^encias, Universidade do Porto, Portugal, Rua das Estrelas, 4150-762 Porto, Portugal}
\author{M.\,S.\,Cunha$^1$}
\author{J.\,Christensen-Dalsgaard}\address{Stellar Astrophysics Centre, Department of Physics and Astronomy, Aarhus University, Ny Munkegade 120, DK-8000 Aarhus C, Denmark}
\begin{abstract}
This work concerns the study of the properties of convective cores 
in main-sequence models of solar-like pulsators and what information they may hold 
about stellar ages. We verified that the maximum absolute frequency derivative 
of particular combinations of frequencies, which we name \lq the slopes\rq, provides information on 
the relative size of the discontinuity in the sound-speed profile at the border of the convectively mixed region. 
Since the latter is related to the evolutionary state of stars, we show that for models with masses above 
$1.3\,\rm M_\odot$, it may be possible to estimate the fraction of stellar main-sequence evolution from the slopes. 
Moreover, for models with masses below $1.2\,\rm M_\odot$ we verified that it may be possible to 
use the slopes to discriminate against models with small amounts of core overshoot. 
\end{abstract}
\maketitle
\section{Introduction}
Asteroseismology is a powerful technique that can be used to 
obtain information about the internal structure of pulsating stars 
through the analysis of their pulsation spectra.

Stars with masses above 1.1\,$\rm M_\odot$ may develop a convective core at some 
stage during their main-sequence evolution. As the star evolves,
the hydrogen abundance is reduced uniformly throughout the convective core
causing an abundance discontinuity at the border of the convectively mixed region. 
This sharp variation causes a discontinuity in the sound-speed profile 
which, in turn, imprints a signature on the star's oscillation frequencies, $\nu_{nl}$. Here, $l$ represents the 
degree of the mode and $n$ its radial order.

The following combination of frequencies was shown to be capable of isolating this signature 
\citep{cunhamet07,cunha11}:
\begin{equation}
\label{eq:dr0213}
dr_{0213} = 6 \left( \frac{D_{02}}{\Delta \nu_{n-1,1}} - \frac{D_{13}}{\Delta \nu_{n,0}}\right),
\end{equation}
where $D_{l,l+2}\equiv(\nu_{n,l}-\nu_{n-1,l+2})/(4l+6)$ and
 $\Delta\nu_{n,l} = \nu_{n+1,l} - \nu_{n,l}$ is the large frequency separation.
Moreover, \cite{cunha11} showed that  the frequency derivative of this diagnostic tool $\Delta\nu dr_{0213}$
can potentially be used to infer the amplitude of the relative sound-speed variation
at the edge of the growing convective core, $A_\delta \equiv [\delta c^2/c^2]_{r=r_d}$, with $r_d$ 
being the radial position at which the discontinuity in the sound speed occurs.

The $l = 3$ modes required to construct the diagnostic tool $dr_{0213}$ may not be always available from observations.
Hence, the following diagnostic tools may be preferred as they also probe 
the inner regions of stars \citep{roxburgh03, gough83},
\begin{equation}
\begin{array}{l}
\label{eq:d0s}
d_{01}(n)=\frac{1}{8}(\nu_{n-1,0}-4\nu_{n-1,1}+6\nu_{n,0}-4\nu_{n,1}+\nu_{n+1,0}), \\ \\
d_{10}(n)=-\frac{1}{8}(\nu_{n-1,1}-4\nu_{n,0}+6\nu_{n,1}-4\nu_{n+1,0}+\nu_{n+1,1}), \\ \\
d_{02}(n)= \nu_{n,0} - \nu_{n-1,2}
\end{array}
\end{equation}
and the respective ratios defined as
\begin{eqnarray}
\label{eq:rat}
r_{01}(n)=\frac{d_{01}(n)}{\Delta \nu_{n,1}},& r_{10} = \frac{d_{10}(n)}{\Delta \nu_{n+1,0}},&  r_{02}(n)=\frac{d_{02}(n)}{\Delta \nu_{n,1}}.
\end{eqnarray}

Note that contrary to $dr_{0213}$, the three diagnostic tools in Eq.\,(\ref{eq:rat})
do not isolate the above mentioned signature but they are still strongly affected by it.
In practice, regarding the two diagnostic tools $d_{01}$ and $d_{10}$, we consider them together, 
denoting the concatenation of the two, appropriately ordered in frequency,
by $d_{010}$, and the corresponding ratios by $r_{010}$.

Since convection implies chemical mixing, 
the evolution of stars harbouring a convective core is greatly 
influenced by the presence and the properties of the convective core. 
In this work we aimed at finding a relation between the frequency
derivatives of the diagnostic tools, computed for a 
set of stellar models considering a large parameter space,
and the relative sound-speed variation at the border of the 
convectively inner region
for the corresponding models. Moreover,
we analysed the potential of this work concerning the inference
of stellar ages.
\section{Method}
\subsection{Stellar models}
\label{sec_stellarmod}
To compute the stellar evolutionary tracks and the model's interior we used the \lq Aarhus STellar Evolution Code\rq\, 
\citep[ASTEC,][]{cd08aastec}. The input parameters used to compute the models are shown in Table\,\ref{tab:inputp} and the input physics 
that we used in this work is described in \cite{brandao11}. We then considered 12 models along the main-sequence phase.
For these models we computed their oscillation frequencies using the Aarhus adiabatic oscillation code \citep[ADIPLS,][]{cd08adipls}.
The computed frequencies were used to calculate the diagnostic tools $dr_{0213}$, $r_{010}$ and $r_{02}$ 
(\textit{cf.} Eq.\,(\ref{eq:dr0213}) and Eq.\,(\ref{eq:rat})) and their frequency derivatives (see Section\,\ref{sec_slopes}).
\begin{table}[htbf]
\centering
\small
\caption{Parameters used to compute the evolutionary tracks. $M/\rm{M_{\odot}}$ is
the mass in solar units, $Z$ is the initial heavy elements abundance, and
$Y$ the helium abundance. $\alpha_{\rm ML}$ is the mixing-length parameter and
$\alpha_{\rm OV}$ the core overshoot parameter.}
\label{tab:inputp}
\begin{tabular}{@{}lll}
\hline\hline
Parameter & Grid I & Grid II\\
\hline
$M/\rm{M_{\odot}}$ & 1.00 - 1.6 (0.1 steps)& 1.00 - 1.6 (0.1 steps)\\
$Z$ & 0.1727 & 0.058, 0.4816\\
$Y$ & 0.278 & 0.255, 0.340\\
$\alpha_{\rm ML}$& 1.8 & 1.8\\
$\alpha_{\rm OV}$& 0.0 - 0.2 (0.1 steps) & 0.1\\
\hline
\end{tabular}
\normalsize
\end{table}
\subsection{Slopes}
\label{sec_slopes}
In the case of stars with a convective core we expect that the
frequency  derivatives of the diagnostic tools taken at their maximum absolute value 
are a measure of $A_\delta$ \citep{cunha11}.  

We introduce here a new quantity $S\{\rm i\}$, which we shall name \lq slope\rq, 
which is a measure of the frequency derivative of a diagnostic tool \lq i\rq, at its maximum absolute value.
The slope is given by the linear coefficient obtained from a linear least-squares fit performed
to the 10 frequencies of modes of consecutive radial orders, $n$, centred on $n_{\rm slope}$.  $n_{\rm slope}$
corresponds to the radial order of the frequency $\nu_{\rm slope}$ at which the maximum absolute value
of the frequency derivative is placed.

Note that, depending on the properties of the models,
the slope may be measured in a frequency range that is 
above the acoustic cut-off frequency, $\nu_{\rm c}$
({\it i.e.}, the maximum frequency such that acoustic waves are expected to be 
contained within the star).

The error associated with
the slope computed for each diagnostic tool
was estimated for some of the models within our grids
with different values of the mass, input
physics, and $A_\delta$. For these
models we randomly generated 10 000 sets of model frequencies 
within the error, assuming a relative error of $10^{-4}$
for each individual frequency. For each generation,
we computed the slopes.
We then computed the mean of the 10 000 values obtained for the slopes
and the standard deviation was considered to be our error
estimation for $S\{\rm i\}$.
\section{Results}
\subsection{Relation between $|S\{i\}|$ and $A_\delta$}
\begin{figure}[!ht]
\begin{center}
  $\begin{array}{l}
  \includegraphics[width=6cm,height=5cm,angle=0]{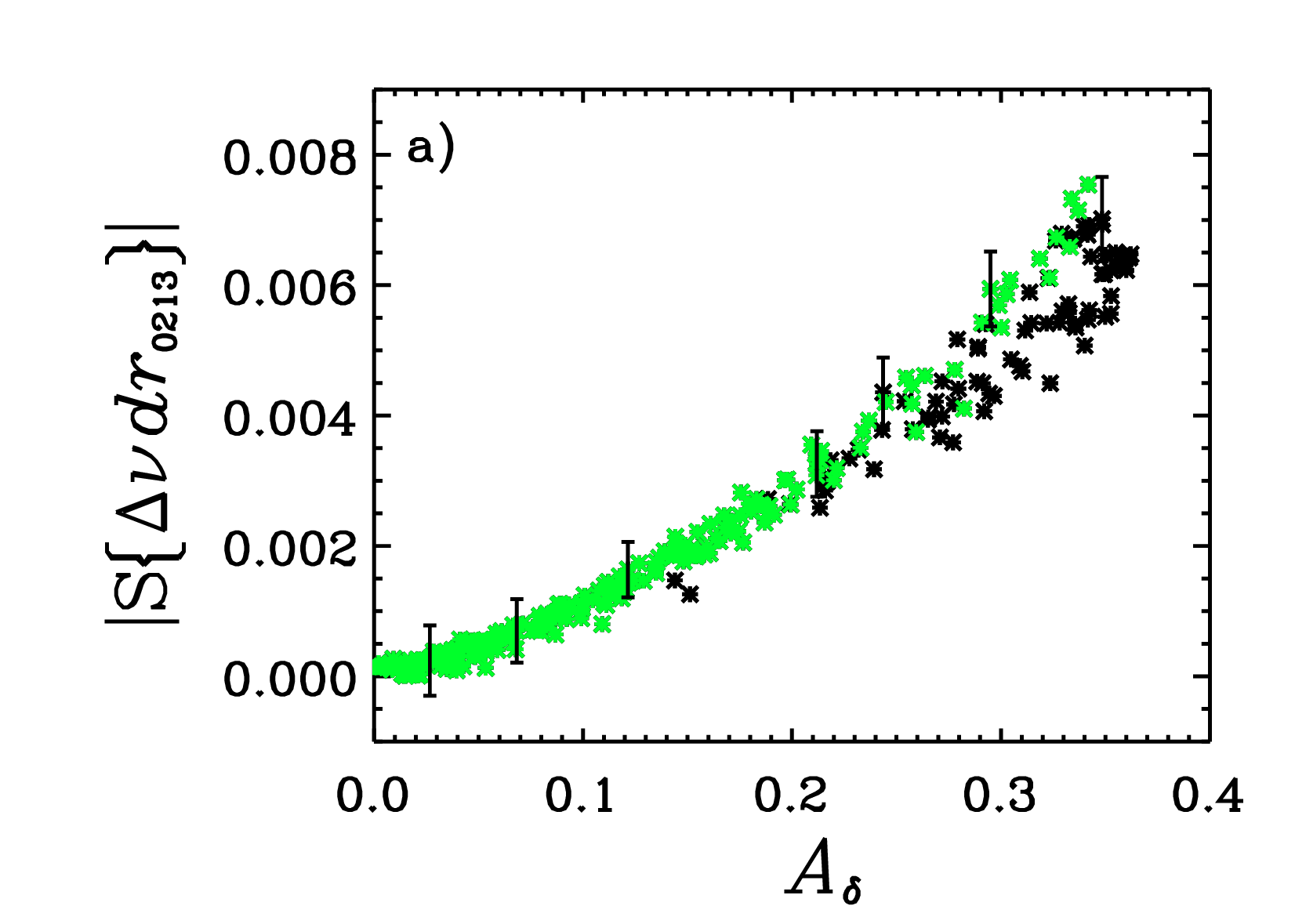}
  \includegraphics[width=6cm,height=5cm,angle=0]{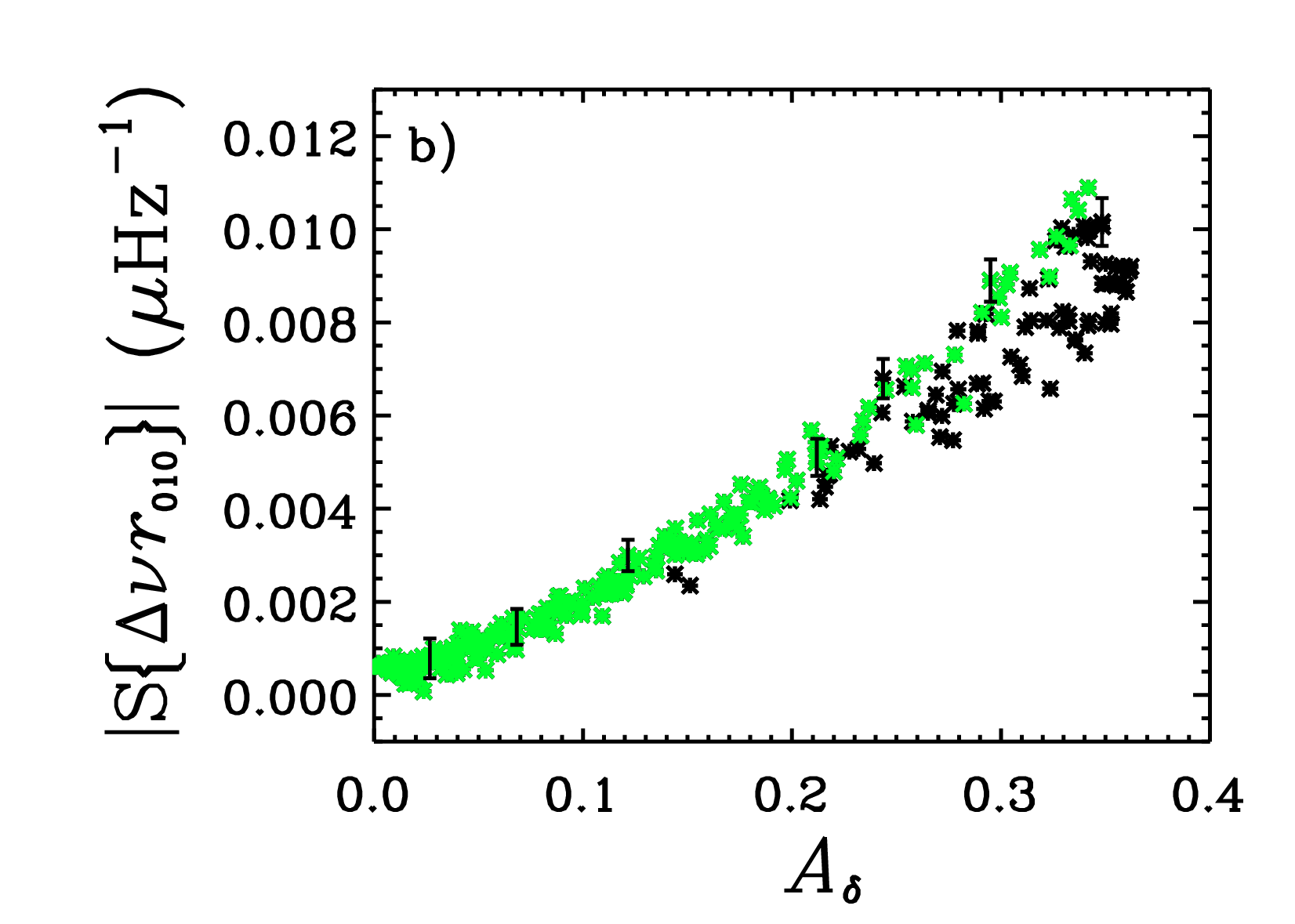} \\
  \includegraphics[width=6cm,height=5cm,angle=0]{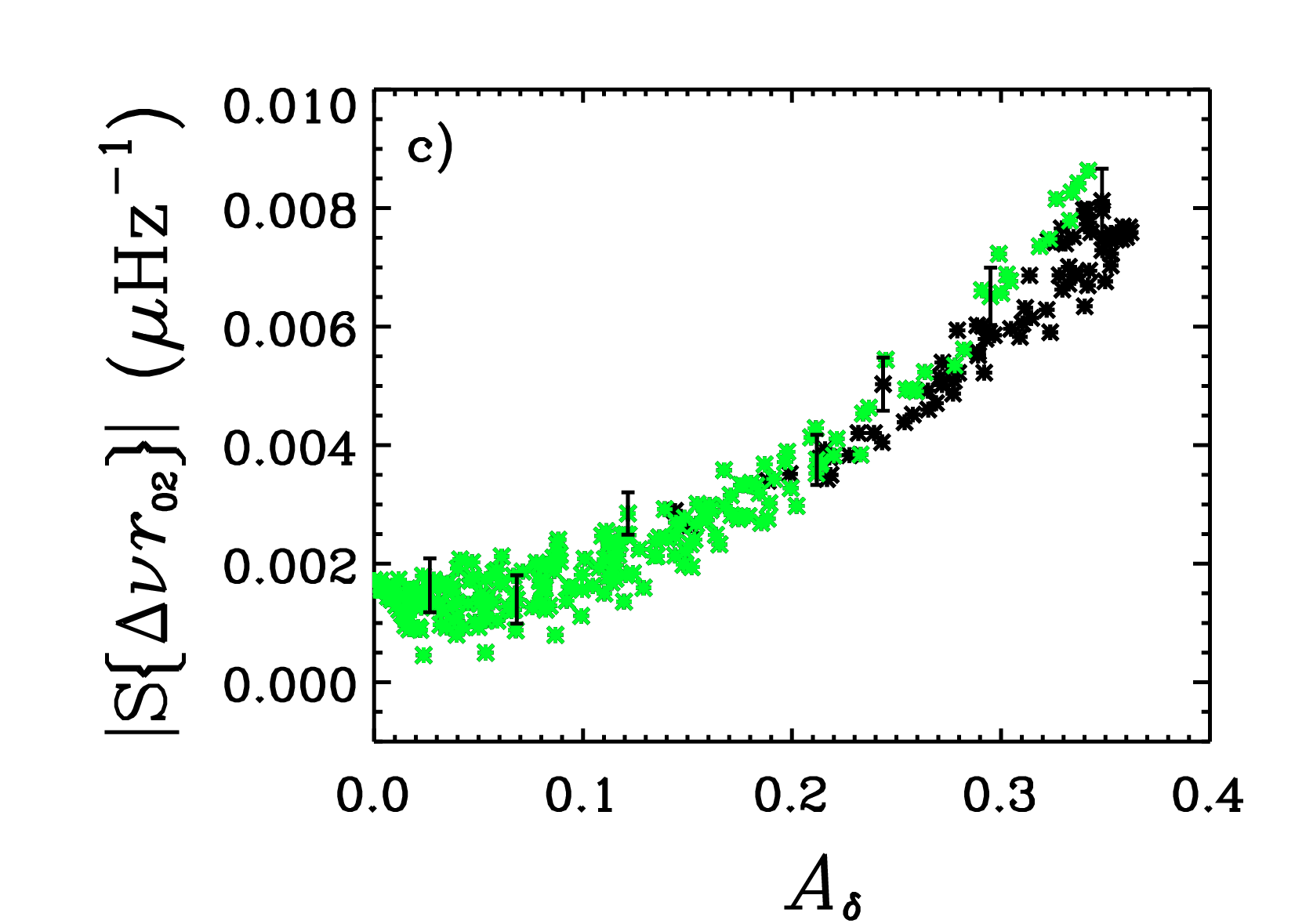} 
  \end{array}$
\end{center}
\caption{The absolute value of the slopes of the diagnostic 
tool $\Delta\nu\,dr_{0213}$ (panel a)), $\Delta\nu\,r_{010}$ (panel b))
and $\Delta\nu\,r_{02}$ (panel c)) as a function of the relative amplitude $A_\delta$ of
the squared sound speed. Each point represented
by a star symbol corresponds to all models in our grid that have a convective core.
From these, the models for which the slope of the diagnostic tools was measured between
$\nu_{\rm max}$ and $\nu_c$ are shown in green.}
\label{fig:slope_deltacc2}
\end{figure}
Fig.\,\ref{fig:slope_deltacc2}\,a), b) and c) show, respectively, 
the relation between the absolute value of the slopes computed 
for the three diagnostic tools $\Delta\nu\,dr_{0213}$,
$\Delta\nu\,r_{010}$, and 
$\Delta\nu\,r_{02}$, and
the relative amplitude $A_\delta$ of the discontinuity 
in the squared sound speed.

As anticipated from the work of \cite{cunha11}, 
there is a strong dependence of $\left| S\{\Delta\nu\,dr_{0213}\} \right|$
on $A_\delta$. Nevertheless, that 
relation deviates from linear and shows significant dispersion, 
particularly for larger values of $A_\delta$. 
In fact, for the younger stars with $A_\delta\lesssim0.2$, the relation
shown in panel a) of Fig.\,\ref{fig:slope_deltacc2} was found not to depend significantly on the mass, core overshooting, 
or metallicity, at least for the physics that we considered in our set of models. 
However, at later stages a dependence of the relation on metallicity and overshoot
emerges as the stars approach the Terminal Age Main Sequence (TAMS). 
The two diagnostic tools $\nu\,r_{010}$ and $\nu\,r_{010}$, unlike $\nu\,dr_{0213}$, 
do not isolate the signature of the sharp structural variation in the sound speed.
Nevertheless, when analysing the slopes of $\Delta\nu\,r_{010}$ and 
$\Delta\nu\,r_{02}$ we obtain results
that are analogous to the ones previously mentioned for
the diagnostic tool $\Delta\nu\,dr_{0213}$.
The dispersion seen in Fig.\,\ref{fig:slope_deltacc2} is significantly
reduced if one considers a subset of models for which the slopes are measured
between the frequency of maximum power, $\nu_{\rm max}$, 
and $\nu_{\rm c}$ (green stars).

Our results demonstrate that it is possible,
for particular stars, to have a measure of $A_\delta$ from the observed frequencies.
\subsection{Relation between $|S\{i\}|$ and $t/t_{\rm TAMS}$}
One would hope that the value derived for $A_\delta$
could be a diagnostic tool for the evolutionary state of the star.
As a test, we inspected directly
the relation between the slopes of the different
diagnostic tools and the star's fraction of evolution ($t/t_{\rm TAMS}$)
along the main-sequence, where $t$ is the age of the star at a given 
evolutionary stage in the main-sequence and $t_{\rm TAMS}$ is the age of the star
within the same evolutionary track but at TAMS. 
\begin{figure}
\begin{center}
\includegraphics[width=6cm,height=4cm,angle=0]{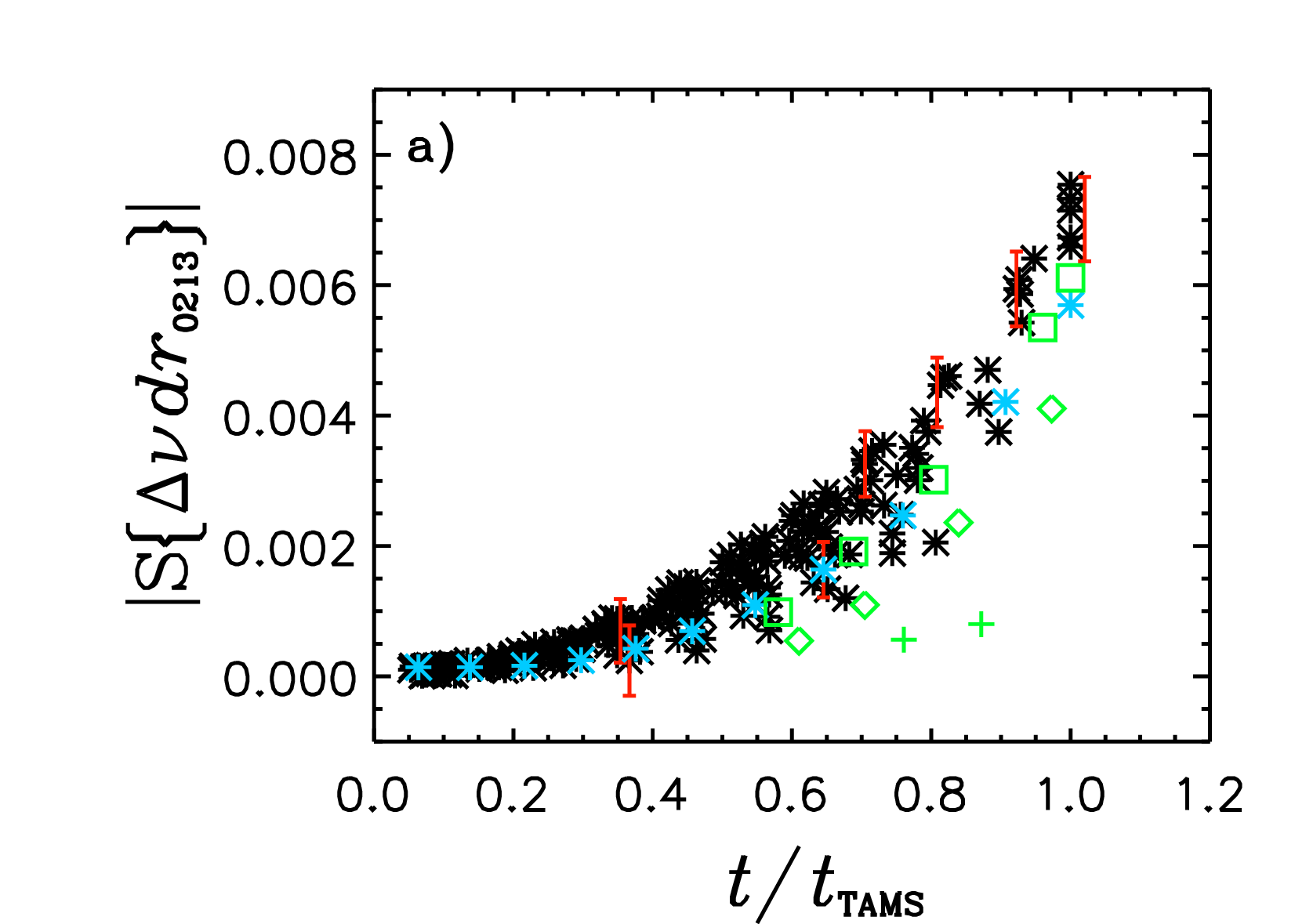}
\includegraphics[width=6cm,height=4cm,angle=0]{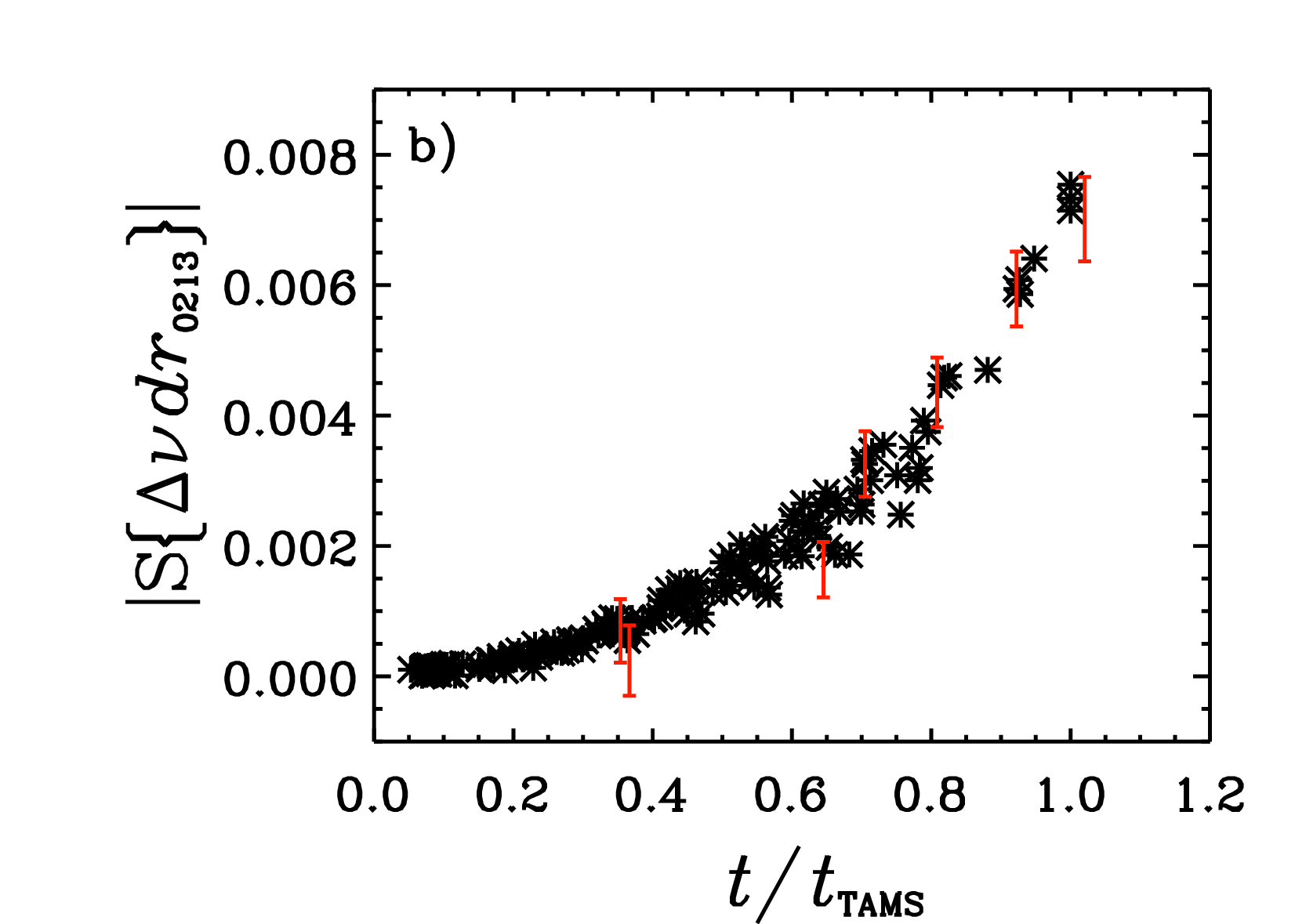} \\
\includegraphics[width=6cm,height=4cm,angle=0]{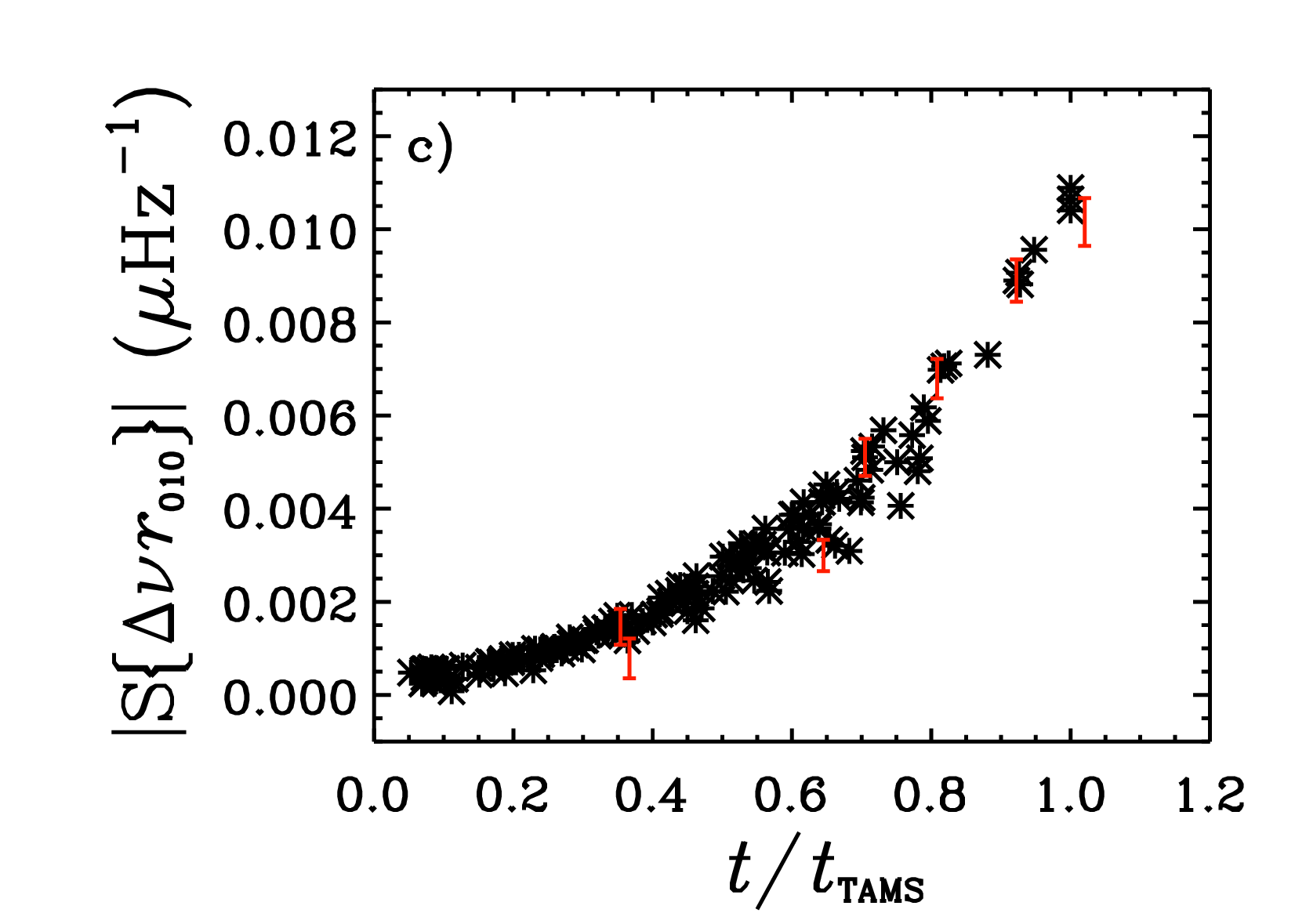} 
\includegraphics[width=6cm,height=4cm,angle=0]{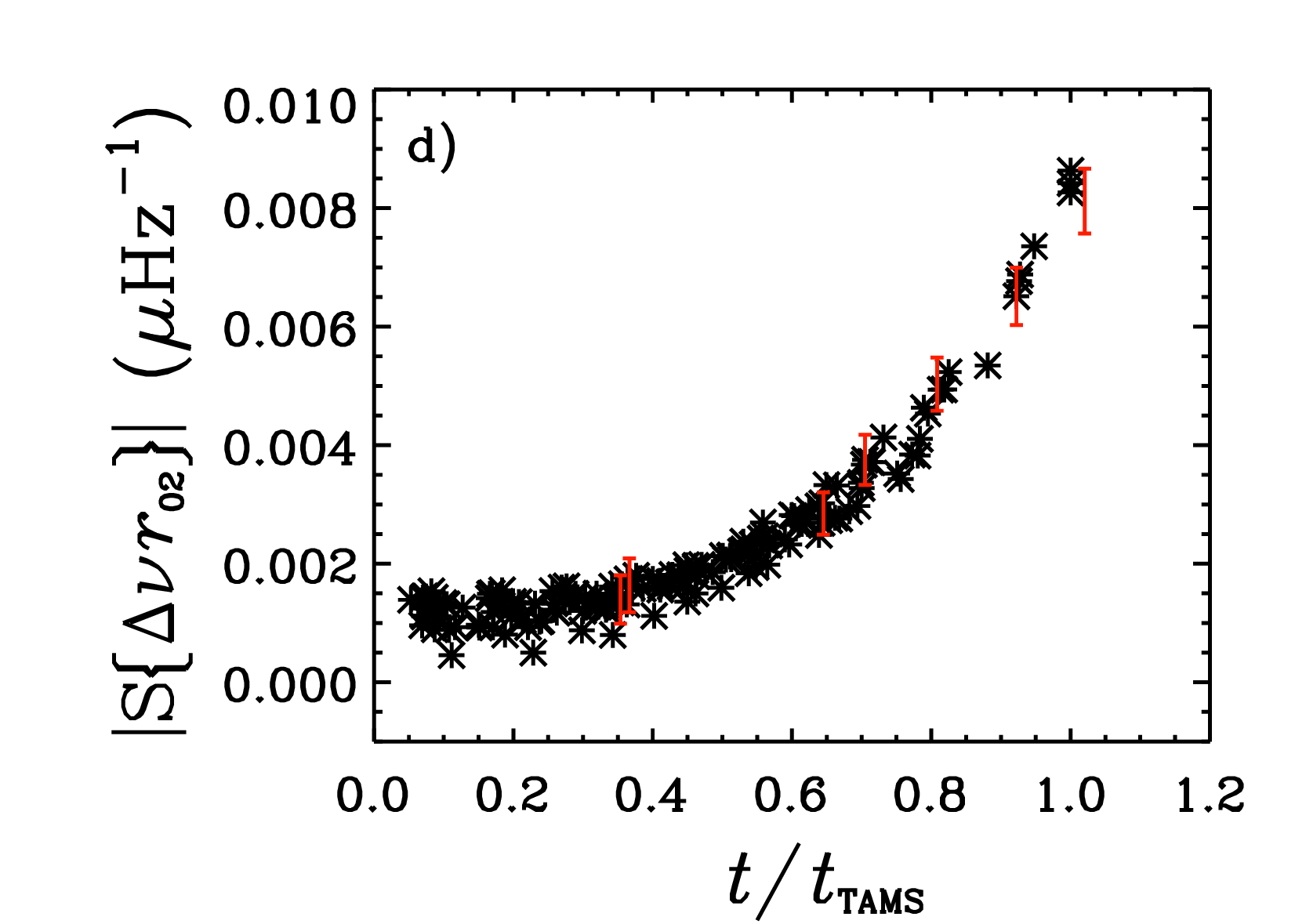} 
\end{center}
\caption{Panel a): The absolute value of the slope of $\Delta\nu dr_{0213}$ 
as a function of the stellar fraction of evolution $t/t_{\rm TAMS}$. Panel b):
The same as in panel a)
but here we only consider models with $M \geq 1.3\,\rm M_\odot$. The same models
are shown in panels c) and d) for the slopes of the 
diagnostic tools $\Delta\nu\,r_{010}$ and $\Delta\nu\,r_{02}$, respectively.}
\label{fig:slope_all_tttams}
\end{figure}

Fig.\,\ref{fig:slope_all_tttams} a) shows, for all
models with a convective core and for which the frequency
region where the slope is computed is between $\nu_{\rm max}$ and $\nu_{\rm c}$,
the absolute value of the slopes of $\Delta\nu\,dr_{0213}$, 
$\left| S\{\Delta\nu\,dr_{0213}\}\right|$, as a function 
of $t/t_{\rm TAMS}$. From this figure, we see a large spread in the slopes 
at $t/t_{\rm TAMS}\gtrsim0.6$.
This is due to the fact that models with $M = 1.1\,\rm M_\odot$,
solar metallicity, and $\alpha_{\rm OV}=0.0$ have a very small central
convective region. As a result, substantial conversion 
of hydrogen into helium takes place 
outside the inner convective region and, in turn, the 
discontinuity in the chemical composition 
at the edge of that convective region is
significantly smaller than in otherwise similar
models with overshoot, at the latest stages of evolution. 

Fig.\,\ref{fig:slope_all_tttams} b)
shows the same plot as in Fig.\,\ref{fig:slope_all_tttams} a)
but considering only those models that have $M \geq 1.3\,\rm M_\odot$.
Here, we see that the dispersion is reduced. 
Fig.\,\ref{fig:slope_all_tttams} c) and d) are the same plots as 
Fig.\,\ref{fig:slope_all_tttams} b) 
but for the diagnostic tools, $\Delta\nu r_{010}$ and $\Delta\nu r_{02}$, respectively.
\section{Conclusions}
We used stellar models 
of different masses, metallicities and convective-core overshoots, 
and at different evolutionary states in the main sequence,
to study he slopes of the diagnostic tools $dr_{0213}$, 
$r_{010}$, and $r_{02}$, where the slopes are defined as 
the frequency derivatives of these quantities taken at their 
maximum absolute value.

In this work we verified that for each evolutionary sequence, 
the absolute value of the slopes increases as the 
star evolves on the main-sequence. This increase 
is associated with the increase in the sound-speed 
discontinuity at the edge of the core. We also verified a strong correlation between the slopes 
and the relative amplitude $A_\delta$ of the discontinuity in the 
squared sound speed.

We observed a relation between the slopes and 
the fraction of stellar 
main-sequence evolution, $t/t_{\rm TAMS}$. This relation is
stronger for the diagnostic tools 
$dr_{0213}$ and $r_{010}$ than for $r_{02}$. 
Also, the dispersion seen in these 
relations is significantly reduced when only models with masses 
$M \geq 1.3\,\rm M_\odot$ are considered. 
We also note that for stars with masses $M \leq 1.2\,\rm M_\odot$, 
the dispersion seen in the slope versus $t/t_{\rm TAMS}$ relation was found 
to be directly related to the amount of core overshoot, 
which in these lower mass stars, with very small convective cores, 
influences the amplitude of the discontinuity in the mean molecular weight
at fixed evolutionary stage. 
As a consequence, for these lower mass stars, it may be possible to 
use the slopes to discriminate against models with small amounts of core overshoot. 
\begin{acknowledgements}
I.M.B. acknowledges the support from the Funda\c{c}\~ao para a Ci\^encia e
Tecnologia (Portugal) through the grant SFRH/BPD/87857/2012.
M.S.C. is supported by an Investigador FCT
contract funded by FCT/MCTES (Portugal) and POPH/FSE (EC). I.M.B. and M.S.C.
acknowledge the support from ERC, under FP7/EC, through the project FP7-SPACE-2012-312844.
Funding for the Stellar Astrophysics Centre is provided by The Danish National Research
Foundation (Grant DNRF106). The research is supported by the ASTERISK project
(ASTERoseismic Investigations with SONG and Kepler) funded by the European Research
Council (Grant agreement no.: 267864).
\end{acknowledgements}
\bibliographystyle{bibtex/aa}
\bibliography{brandao_im.bib} 
\end{document}